\documentclass[italian,english]{article}
\usepackage[latin9]{inputenc}
\usepackage{amstext}
\usepackage{amssymb}

\makeatletter
\newcommand{\lyxaddress}[1]{
\par {\raggedright #1
\vspace{1.4em}
\noindent\par}
}

\makeatother

\usepackage{babel}

\begin{document}

\title{\textbf{High-energy scalarons in $R^{2}$ gravity as a model for
Dark Matter in galaxies}}

\author{\textbf{C. Corda$^{1},$ H. J. Mosquera Cuesta$^{1,2,3,4},$ R. Lorduy
Gòmez$^{5}$ }}

\maketitle

\lyxaddress{\begin{center}
\textbf{$^{1}$}International Institute for Theoretical Physics and
Mathematics Einstein-Galilei, Via Santa Gonda, 14 - 59100 Prato, Italy%
\footnote{\begin{quotation}
CC is partially supported by a Research Grant of The R. M. Santilli
Foundation Number RMS-TH-5735A2310
\end{quotation}
}
\par\end{center}}

\lyxaddress{\begin{center}
\textbf{$^{2}$}Departmento de Fìsica Universidade Estadual Vale
do Acaraù, Avenida da Universidade 850, Campus da Betània, CEP 62.040-370,
Sobral, Cearà, Brazil%
\footnote{HJMC is fellow of the Cearà State Foundation for the Development of
Science and Technology (FUNCAP), Fortaleza, CE, Brazil%
}
\par\end{center}}

\lyxaddress{\begin{center}
$^{3}$Instituto de Cosmologia, Relatividade e Astrofìsica (ICRA-BR),
Centro Brasilero de Pesquisas Fìsicas, Rua Dr. Xavier Sigaud 150,
CEP 22290 -180 Urca Rio de Janeiro, RJ, Brazil
\par\end{center}}

\begin{center}
$^{4}$International Center for Relativistic Astrophysics Network
(ICRANet), International Coordinating Center, Piazzale della Repubblica
10, 065112, Pescara, Italy
\par\end{center}

\lyxaddress{\begin{center}
\textbf{$^{5}$}Grupo de Astronomìa Quasar, Universidad EAFIT, Medellìn,
Colombia
\par\end{center}}
\begin{abstract}
We show that in the framework of $\textbf{\ensuremath{R^{2}}}$ gravity
and in the linearized approach it is possible to obtain spherically
symmetric stationary states that can be used as a model for galaxies.
Such approach could represent a solution to the Dark Matter Problem.
In fact, in the model, the Ricci curvature generates a high energy
term that can in principle be identified as the dark matter field
making up the galaxy. The model can also help to have a better understanding
on the theoretical basis of Einstein-Vlasov systems. Specifically,
we discuss, in the linearized $\textbf{\ensuremath{R^{2}}}$ gravity,
the solutions of a Klein-Gordon equation for the spacetime curvature.
Such solutions describe high energy \textit{scalarons}, a field that
in the context of galactic dynamics can be interpreted like the no-light-emitting
galactic component. That is, these particles can be figured out like
wave-packets showing stationary solutions in the Einstein-Vlasov system.
In such approximation, the energy of the particles can be thought
of as the galactic dark matter component that guarantees the galaxy
equilibrium. Thus, because of the high energy of such particles the
coupling constant of the $\textbf{\ensuremath{R^{2}}}$-term in the
gravitational action comes to be very small with respect to the linear
term $R$. In this way, the deviation from standard General Relativity
is very weak, and in principle the theory could pass the Solar System
tests. As pertinent to the issue under analysis in this paper, we
present an analysis on the gravitational lensing phenomena within
this framework. 

Although the main goal of this paper is to give a potential solution
to the Dark Matter Problem within galaxies, we add a Section where
we show that an important property of the Bullet Cluster can in principle
be explained in the scenario introduced in this work. 

To the end, we discuss the generic prospective to give rise to the
Dark Matter component of most galaxies within extended gravity. 
\end{abstract}

\lyxaddress{PACS numbers: 95.35.+d, 04.50.Kd.}


\section{Introduction}

The accelerated expansion of the universe, that is today observed,
suggests that cosmological dynamics is dominated by the so-called
Dark Energy field which provides a large negative pressure. This is
the standard picture, in which such new ingredient is considered as
a source of the \textit{right hand side} of the field equations. It
should be some form of non-clustered non-zero vacuum energy which,
together with the clustered Dark Matter, drives the global dynamics.
This is the so-called {}``\emph{concordance model}'' ($\Lambda$CDM)
which gives, in agreement with the Cosmic Microwave Background Radiation
(CMBR), dim Lyman Limit Systems (LLS) and type la supernovae (SNeIa)
data, a good framework to understand the today observed Universe.
However, it presents several shortcomings as the well known ''\emph{coincidence}''
and \emph{{}``cosmological constant}'' problems \cite{key-1}. An
alternative approach is to change the \emph{left hand side} of the
field equations, and check if observed cosmic dynamics can be achieved
by extending general relativity \cite{key-2,key-3,key-4,key-5,key-6}.
In this different context, it is not required to search candidates
for Dark Energy and Dark Matter, which till now have not been found.
Rather, one can only stand on the \emph{{}``observed}'' ingredients:
curvature and baryon matter, to account for the observations. Considering
this point of view, one can think of that gravity is not scale-invariant.
Such an assumption opens a room for alternative theories to be introduced
\cite{key-7,key-8,key-9}. In principle, the most popular Dark Energy
and Dark Matter models can be achieved by considering $f(R)$ theories
of gravity \cite{key-2}-\cite{key-9}, where $R$ is the Ricci curvature
scalar. 

In this picture, even the sensitive detectors for gravitational waves
like bars and interferometers (i.e. those which are currently in operation
and the ones which are in a phase of planning and proposal stages)
\cite{key-10} could, in principle, be important to confirm or rule
out the physical consistency of general relativity or of any other
theory of gravitation. This is because, in the context of Extended
Theories of Gravity, some differences between General Relativity and
the alternative theories can be pointed out as far as the linearized
theory of gravity is concerned \cite{key-11,key-12}.

In the general picture of high order theories of gravity, recently
the $\textbf{\ensuremath{R^{2}}}$ theory, which was originally proposed
by Starobinski \cite{key-13}, has been analysed in various interesting
frameworks, see \cite{key-14,key-15,key-16,key-17} for example. Specifically,
the non-singular behaviour of this class of models is discussed in
\cite{key-14}. In \cite{key-15} $\textbf{\ensuremath{R^{2}}}$ inflation
is combined with the Dark Energy stage and in \cite{key-16} an oscillating
Universe, which is well tuned with some cosmological observations
is discussed. Finally, in \cite{key-17} the production and potential
detection of gravity-waves from this particular theory has been shown.

It is also quite important to emphasize that the $\textbf{\ensuremath{R^{2}}}$
is the simplest one among the class of viable models with $R^{m}$
terms in addition to the Einstein-Hilbert theory. In Ref. \cite{key-5},
it has been shown that such models may lead to the (cosmological constant
or quintessence) acceleration of the universe as well as an early
time era of inflation. Moreover, they seem to pass the Solar System
tests, i.e. they have the acceptable newtonian limit, no instabilities
and no Brans-Dicke problem (decoupling of the scalar) in the scalar-tensor
version. 

In this paper it is shown that in the framework of $\textbf{\ensuremath{R^{2}}}$
gravity, in the linearized approach, it is possible to obtain spherically
symmetric and stationary galaxy states which can be interpreted like
an approximated solution to the Dark Matter Problem. In fact, in the
proposed model the Ricci curvature scalar generates an energy term
that can in principle be identified as the Dark Matter content in
the galaxy. The model can also help to have a better understanding
of the physics of Einstein-Vlasov systems \cite{key-18}. 

As the observed gravitational lensing phenomena are well related to
the S-type galaxies in the Hubble sequence, here we briefly discuss
in Section 4 the implications of this theory for observations of bending
of light by foreground galaxies. 

The main goal of this paper is to give a potential solution to the
Dark Matter Problem within galaxies. In other words, we do not ban
the presence of a Dark Matter component in our Universe which can
explain properties of clusters of galaxies which cannot be achieved
by alternative gravity theories. In any case, we add a Section where
we show that an important property of the bullet cluster can be, in
principle, explained in the scenario given in this paper \cite{key-44}.

At the end of the paper, we discuss the general possibility to give
rise to a Dark Matter within extended gravity.

\section{The field equations and the linearized approach}

Let us consider the high order action \cite{key-15,key-16,key-17}

\begin{equation}
S=\int d^{4}x\sqrt{-g}(R+bR^{2}+\mathcal{L}_{m}).\label{eq: high order 1}\end{equation}
In order to avoid confusion, in all this paper $b$ represents the
coupling constant of the $R^{2}$ term. 

The action (\ref{eq: high order 1}) is a particular choice with respect
to the well known canonical action characterizing General Relativity
(the Einstein - Hilbert action \cite{key-19}), which reads 

\begin{equation}
S=\int d^{4}x\sqrt{-g}(R+\mathcal{L}_{m}).\label{eq: EH}\end{equation}

If the gravitational Lagrangian is not linear in the curvature invariants,
the Einstein field equations has an order higher than second. For
this reason, such theories are often called higher-order gravitational
theories \cite{key-2}-\cite{key-9}. This is exactly the case of
the action (\ref{eq: high order 1}). Note that in this paper we work
with $8\pi G=1$, $c=1$ and $\hbar=1,$ while the sign conventions
for the line element, which generate the sign conventions for the
Riemann/Ricci tensors, are $(-,+,+,+)$. 

By varying the action (\ref{eq: high order 1}) with respect to $g_{\mu\nu}$
(see refs. \cite{key-11,key-16,key-17} for a parallel computation)
the field equations are obtained:

\begin{equation}
\begin{array}{c}
G_{\mu\nu}+b\{2R[R_{\mu\nu}-\frac{1}{4}g_{\mu\nu}R]+\\
\\-2R_{;\mu;\nu}+2g_{\mu\nu}\square R\}=T_{\mu\nu}^{(m)}\end{array}\;,\label{eq: einstein-general}\end{equation}

with the associated Klein - Gordon equation for the Ricci curvature
scalar 

\begin{equation}
\square R=E^{2}(R+T),\label{eq: KG}\end{equation}

which is obtained by taking the trace of equation (\ref{eq: einstein-general}),
where $\square$ is the d'Alembertian operator and the energy term,
$E$, has been introduced for dimensional motivations:

\begin{equation}
E^{2}\equiv\frac{1}{6b}\;,\label{eq: Eya}\end{equation}

thus, $b$ has to be positive \cite{key-16}.

In the above equations $T_{\mu\nu}^{(m)}$ is the standard stress-energy
tensor of the matter. Note that General Relativity is obtained for
$b=0$ in Eq. (3).

Because we want to study interactions between stars at galactic scales,
the linearized theory in vacuum ($T_{\mu\nu}^{(m)}=0$), which gives
a better approximation than Newtonian theory, can be analysed, by
considering a small perturbation of the background, which is assumed
to be given by a Minkowskian background. 

In the linearization procedure we will perform a computation very
similar to the ones in \cite{key-16,key-17}, but with some difference
that will emphasize the importance of the Ricci scalar for the purpose
of this paper: to provide a curvature-inspired model for galaxies.

Putting

\begin{equation}
g_{\mu\nu}=\eta_{\mu\nu}+h_{\mu\nu}\label{eq: linearizza}\end{equation}

to first order in $h_{\mu\nu}$, calling $\widetilde{R}_{\mu\nu\rho\sigma}$,
$\widetilde{R}_{\mu\nu}$ and $\widetilde{R}$, respectively, the
linearized quantity which correspond to $R_{\mu\nu\rho\sigma}$, $R_{\mu\nu}$
and $R$, the linearized field equations are obtained as \cite{key-16,key-17}:

\begin{equation}
\begin{array}{c}
\widetilde{R}_{\mu\nu}-\frac{\widetilde{R}}{2}\eta_{\mu\nu}=-\partial_{\mu}\partial_{\nu}b\widetilde{R}+\eta_{\mu\nu}\square b\widetilde{R}\\
\\{}\square\widetilde{R}=E^{2}\widetilde{R}.\end{array}\label{eq: linearizzate1}\end{equation}

Since $\widetilde{R}_{\mu\nu\rho\sigma}$ and Eqs. (\ref{eq: linearizzate1})
are invariants for gauge transformations of the kind \cite{key-16,key-17}

\begin{equation}
\begin{array}{c}
h_{\mu\nu}\rightarrow h'_{\mu\nu}=h_{\mu\nu}-\partial_{(\mu}\epsilon_{\nu)}\\
\\b\widetilde{R}\rightarrow b\widetilde{R}'=b\widetilde{R};\end{array}\label{eq: gauge}\end{equation}

then the gauge transformation

\begin{equation}
\bar{h}_{\mu\nu}\equiv h_{\mu\nu}-\frac{h}{2}\eta_{\mu\nu}+\eta_{\mu\nu}b\widetilde{R}\label{eq: ridefiniz}\end{equation}

can be defined. Thus, by considering the transformation rule for the
parameter $\epsilon^{\mu}$

\begin{equation}
\square\epsilon_{\nu}=\partial^{\mu}\bar{h}_{\mu\nu},\label{eq:lorentziana}\end{equation}

a gauge analogous to the Lorenz one for electromagnetic waves can
be chosen, which then reads

\begin{equation}
\partial^{\mu}\bar{h}_{\mu\nu}=0.\label{eq: cond lorentz}\end{equation}

In this way, the field equations then read like

\begin{equation}
\square\bar{h}_{\mu\nu}=0,\label{eq: onda T}\end{equation}

\begin{equation}
\square b\widetilde{R}=E^{2}b\widetilde{R}.\label{eq: onda S}\end{equation}

Solutions of eqs. (\ref{eq: onda T}) and (\ref{eq: onda S}) are
plan waves:

\begin{equation}
\bar{h}_{\mu\nu}=A_{\mu\nu}(\overrightarrow{p})\exp(ip^{\beta}x_{\beta})+c.c.\label{eq: sol T}\end{equation}

\begin{equation}
b\widetilde{R}=a(\overrightarrow{p})\exp(iq^{\beta}x_{\beta})+c.c.\label{eq: sol S}\end{equation}

where

\begin{equation}
\begin{array}{ccc}
p^{\beta}\equiv(\omega,\overrightarrow{p}), &  & \omega=p\equiv|\overrightarrow{p}|\\
\\q^{\beta}\equiv(\omega_{E},\overrightarrow{p}), &  & \omega_{E}=\sqrt{E^{2}+p^{2}}.\end{array}\label{eq: k e q}\end{equation}

In Eqs. (\ref{eq: onda T}) and (\ref{eq: sol T}) the equation and
the solution for the tensor waves, exactly like in general relativity,
have been obtained \cite{key-16,key-17}, while Eqs. (\ref{eq: onda S})
and (\ref{eq: sol S}) are, respectively, the equation and the solution
for the third mode stemming from the curvature (see also \cite{key-16,key-17}).

The fact that the dispersion law for the modes of the field $b\widetilde{R}$
is not linear has to be emphasized. The velocity of every {}``\emph{ordinary}''
(i.e. which arises from General Relativity) mode $\bar{h}_{\mu\nu}$
is the light speed $c$, but the dispersion law (the second of Eqs.
(\ref{eq: k e q})) for the modes of $b\widetilde{R}$ is that of
a wave-packet \cite{key-16,key-17}. Also, the group-velocity of a
wave-packet of $b\widetilde{R}$ centred in $\overrightarrow{p}$
is 

\begin{equation}
\overrightarrow{v_{G}}=\frac{\overrightarrow{p}}{\omega}.\label{eq: velocita' di gruppo}\end{equation}

From the second of Eqs. (\ref{eq: k e q}) and Eq. (\ref{eq: velocita' di gruppo})
it is simple to obtain:

\begin{equation}
v_{G}=\frac{\sqrt{\omega^{2}-E^{2}}}{\omega}.\label{eq: velocita' di gruppo 2}\end{equation}

Then, it is also written like \cite{key-16,key-17}

\begin{equation}
E=\sqrt{(1-v_{G}^{2})}\omega.\label{eq: relazione massa-frequenza}\end{equation}

Now, the analysis can remain in the Lorenz gauge with transformations
of the type $\square\epsilon_{\nu}=0$. This gauge gives a condition
of transverse effect for the ordinary part of the field, i.e. $k^{\mu}A_{\mu\nu}=0$,
but does not give the transverse effect for the total field $h_{\mu\nu}$.

From Eq. (\ref{eq: ridefiniz}) it reads

\begin{equation}
h_{\mu\nu}=\bar{h}_{\mu\nu}-\frac{\bar{h}}{2}\eta_{\mu\nu}+\eta_{\mu\nu}b\widetilde{R}.\label{eq: ridefiniz 2}\end{equation}

At this point, if one were still to remain within general relativity
one would impose \cite{key-23},

\begin{equation}
\begin{array}{c}
\square\epsilon^{\mu}=0\\
\\\partial_{\mu}\epsilon^{\mu}=-\frac{\bar{h}}{2}+b\widetilde{R},\end{array}\label{eq: gauge2}\end{equation}

which would give the total transverse effect of the field. However,
for the present case this is impossible. In fact, applying the d'Alembertian
operator to the second of Eqs. (\ref{eq: gauge2}) and using the field
equations (\ref{eq: onda T}) and (\ref{eq: onda S}), it comes out
that

\begin{equation}
\partial_{\mu}\square\epsilon^{\mu}=E^{2}b\widetilde{R},\label{eq: contrasto}\end{equation}

which is in contrast with the first of Eqs. (\ref{eq: gauge2}). In
the same way, it is possible to show that no linear relation exists
at all between the tensor field $\bar{h}_{\mu\nu}$ and the linearized
term $b\widetilde{R}$ stemming from the curvature scalar. Thus, a
gauge in which $h_{\mu\nu}$ is purely spatial cannot be chosen (i.e.
one cannot put $h_{\mu0}=0,$ see Eq. (\ref{eq: ridefiniz 2})). 

Nonetheless, the traceless condition to the field $\bar{h}_{\mu\nu}$
can be written in the form:

\begin{equation}
\begin{array}{c}
\square\epsilon^{\mu}=0\\
\\\partial_{\mu}\epsilon^{\mu}=-\frac{\bar{h}}{2}.\end{array}\label{eq: gauge traceless}\end{equation}

These equations imply

\begin{equation}
\partial^{\mu}\bar{h}_{\mu\nu}=0.\label{eq: vincolo}\end{equation}

To save the conditions $\partial_{\mu}\bar{h}^{\mu\nu}=0$ and $\bar{h}=0,$
transformations like

\begin{equation}
\begin{array}{c}
\partial_{\mu}\square\epsilon^{\mu}=0\\
\\\partial_{\mu}\epsilon^{\mu}=0\end{array}\label{eq: gauge 3}\end{equation}

can be used and, by taking $\overrightarrow{p}$ in the $z$ direction,
a gauge in which only $A_{11}$, $A_{22}$, and $A_{12}=A_{21}$ are
different from zero can be chosen. The condition $\bar{h}=0$ gives
$A_{11}=-A_{22}$. 

Now, by substituting these equations in Eq. (\ref{eq: ridefiniz 2}),
one gets

\begin{equation}
h_{\mu\nu}(t,z)=A^{+}(t-z)e_{\mu\nu}^{(+)}+A^{\times}(t-z)e_{\mu\nu}^{(\times)}+b\widetilde{R}(t,z)\eta_{\mu\nu}.\label{eq: perturbazione totale}\end{equation}

The term $A^{+}(t-z)e_{\mu\nu}^{(+)}+A^{\times}(t-z)e_{\mu\nu}^{(\times)}$
describes the two standard polarizations of gravitational waves obtained
from general relativity, while the term $b\widetilde{R}(t,z)\eta_{\mu\nu}$
represents a mode arising from the curvature term associate to the
$\textbf{\ensuremath{R^{2}}}$ high order theory \cite{key-16,key-17}. 

In other words, in the $\textbf{\ensuremath{R^{2}}}$ theory of gravity,
the linearized Ricci scalar is a source of a third polarization mode
for gravitational waves which is not present in standard general relativity.
This third mode is associated to a \emph{{}``curvature}'' energy
$E$ (see Eq. (\ref{eq: KG})).

We also recall that the idea of considering the Ricci scalar as an
effective scalar field (\emph{scalaron}) arose from Starobinski \cite{key-13}.

\section{Application to the Einstein-Vlasov system: stationary galaxies }

Now, we are going to discuss a model of stationary, spherically symmetric
galaxy, assuming that the dynamics of the matter, i.e. of the stars
making of the galaxy, is described by the Enstein-Vlasov system \cite{key-18}.
In this way, the gravitational forces between the particles of the
system, i.e., a galaxy, will be mediated by the third mode of Eq.
(\ref{eq: perturbazione totale}), i.e. by the spacetime curvature.
Thus, the key assumption is that in a cosmological context such a
mode, which is given by the (linearized) spacetime curvature, becomes
dominant at galactic and cosmological scales (i.e. $A^{+},A^{-}\ll b\widetilde{R}$).
In this way the {}``\emph{curvature}'' energy $E$ can be identified
as the dark matter content of a galaxy of typical mass-energy: $E\simeq10^{45}g$
\cite{key-25}, in ordinary c.g.s. units. These two assumptions constitute
an adaptation to modelling a galaxy of the assumptions related to
the model describing an oscillating Universe in \cite{key-16}. 

Now, let us emphasize an important point. Assuming $E\simeq10^{45}g,$
from Eq. (\ref{eq: Eya}) we get $b\simeq10^{-34}cm^{4}$ in natural
units. Thus, in our assumption, the constant coupling of the $\textbf{\ensuremath{R^{2}}}$
term in the gravitational action results infinitesimal with respect
to the linear term $R\mbox{\mbox{. }}$ In this way, the variation
from standard General Relativity is very weak, thus the theory can
pass the Solar System tests. Regarding this important issue, it is
important to provide citations of precedent work illustrating this
and explicitly show that the bounds there agree with the preferred
values of the number $E$. The key point is that as the effective
scalar field arising from curvature is very energetic, then the constant
coupling of the the $\textbf{\ensuremath{R^{2}}}$ nonlinear term
$\rightarrow0$ \cite{key-31}. In this case, the Ricci curvature,
which is an extra dynamical quantity in the metric formalism, must
have a range longer than the size of the Solar System. An important
work is ref. \cite{key-32}, where it is shown that this is correct
if the effective length of the scalar field $l$ is much shorter than
the value of $0.2$ $mm$. In such a case, the presence of this effective
scalar is hidden from Solar System and terrestrial experiments. The
value of $E$ that we are assuming here guarantees the condition $l\ll0.2$
$mm$. Another important test concerns the deflection of light by
the Sun. This effect was studied in $\textbf{\ensuremath{R^{2}}}$
gravity by calculating the Feynman amplitudes for photon scattering,
and it was found that, to linearized order, this deflection is the
same as in standard General Relativity \cite{key-33}. 

On the other hand, Eq. (\ref{eq: sol S}) guarantees that spacetime
curvature does not remain too small and thus it can, in principle,
become the Dark Matter component in the galaxy \cite{key-24}.

The model that we shall discuss is similar to the one introduced by
Nordstrom in \cite{key-21}. The relevance of the Einstein-Vlasov
model was emphasized in the 1992 in a famous paper by Rein and Rendall
\cite{key-18}. And the following results will be obtained adapting
the ideas introduced in \cite{key-18,key-22,key-24,key-26,key-28}.

In the hypothesis $A^{+},A^{-}\ll b\widetilde{R}$, the spacetime
of our model will be given by the conformally flat metric

\begin{equation}
ds^{2}=[1+b\widetilde{R}(t,z)](dx^{2}+dy^{2}+dz^{2}-dt^{2}).\label{eq: metrica puramente scalare}\end{equation}

Note: in general, conformal transformations are performed by rescaling
the line-element like \cite{key-23}

\begin{equation}
\tilde{g}_{\alpha\beta}=e^{\Phi}g_{\alpha\beta}.\label{eq: conforme}\end{equation}

Here we choose the scalar field as being

\begin{equation}
\Phi\equiv b\widetilde{R}\label{eq: trucco conforme 2}\end{equation}

which also implies 

\begin{equation}
e^{\Phi}=1+b\widetilde{R},\label{eq: trucco conforme}\end{equation}

in our linearized approach. Thus, the conformal transformation which
translates the analysis into the conformal frame (Einstein frame,
see \cite{key-5}) is performed by the spacetime curvature. The potential
of using such conformally flat line-element in the perspective of
describing an oscillating Universe has been discussed in \cite{key-16}.

The condition that the particles in the spacetime make up an ensemble
with no collisions (\ref{eq: metrica puramente scalare}) is satisfied
if the particle density is a solution of the Vlasov equation \cite{key-18,key-22,key-26,key-28}

\begin{equation}
\partial_{t}f+\frac{p^{a}}{p^{0}}\partial_{x^{a}}f-\Gamma_{\mu\nu}^{a}\frac{p^{\mu}p^{\nu}}{p^{0}}\partial_{p^{a}}f=0,\label{eq: Vlasov}\end{equation}

where $\Gamma_{\mu\nu}^{\alpha}$ are the Christoffel coefficients,
$f$ is the particle density and $p^{0}$ is given by $p^{a}$($a=1,2,3$)
according to the relation \cite{key-18,key-22,key-26,key-28}

\begin{equation}
g_{\mu\nu}p^{\mu}p^{\nu}=-1.\label{eq: mass-shell}\end{equation}

Eq. (\ref{eq: mass-shell}) implies that the four momentum $p^{\mu}$
lies on the mass-shell of the spacetime (greek indices run from 0
to 3) \cite{key-18,key-22,key-26,key-28}. 

We recall that, in general, the Vlasov-Poisson system is given by
\cite{key-18,key-22,key-26,key-28}

\begin{equation}
\begin{array}{c}
\partial_{t}f+v\cdot\bigtriangledown_{x}f-\bigtriangledown_{x}U\cdot\bigtriangledown_{v}f=0\\
\\\bigtriangledown\centerdot U=4\pi\rho\\
\\\rho(t,x)=\int dvf(t,x,v),\end{array}\label{eq: VP}\end{equation}

where $t$ denotes the time and $x$ and $v$ the position and the
velocity of the stars. The function $U=U(t,x)$ is the average Newtonian
potential generated by the stars. This system represents the non-relativistic
kinetic model for an ensemble of particles with no collisions interacting
through gravitational forces which they generate collectively \cite{key-18,key-22,key-26,key-28}. 

Thus, such a system can be used for a description of the motion of
the stars within a galaxy, if stars are considered as pointlike particles,
and the relativistic effects are negligible \cite{key-18,key-22,key-26,key-28}.
In this approach, the function $f(t,x,v)$ in the Vlasov-Poisson system
(\ref{eq: VP}) is non-negative and gives the density on phase space
of the stars within the galaxy.

In other words, we are going to discuss the solutions of a Klein-Gordon
equation for the spacetime curvature like galactic high energy \emph{scalarons},
i.e. particles that can be figured out like wave-packets, and stationary
solutions in terms of an Einstein-Vlasov system will be shown. 

In such approximation, the energy of the particle will be seen like
the Dark Matter component that guarantees the galaxy's equilibrium.
It is stressed that such approximation is not as precise as we would,
but it could be a starting point for further robust analysis.

The Vlasov equation (\ref{eq: Vlasov}) implies that the function
$f$ is constant on the geodesic flow over the spacetime (\ref{eq: metrica puramente scalare}).
The Christoffel coefficients of such a spacetime are obtained from
(note that as we are working in the linearized approach, in the following
computations only terms up to first order in the linearized curvature
$\widetilde{R}$ will be considered while high-order terms will be
assumed equal to zero)

\begin{equation}
\Gamma_{\mu\nu}^{\alpha}=\frac{1}{2}(\delta_{\nu}^{\alpha}\partial_{\mu}b\widetilde{R}+\delta_{\mu}^{\alpha}\partial_{\nu}b\widetilde{R}-\frac{1}{1+2b\widetilde{R}}g_{\mu\nu}\partial^{\alpha}b\widetilde{R}).\label{eq: connessioni}\end{equation}

In this way, the Vlasov equation in the spacetime defined by the line-element
(\ref{eq: metrica puramente scalare}) becomes

\begin{equation}
\partial_{t}f+\frac{p^{a}}{p^{0}}\partial_{x^{a}}f-\frac{1}{2}\left[2(p^{\mu}\partial_{\mu}b\widetilde{R})\frac{p^{a}}{p^{0}}+\frac{\partial^{a}b\widetilde{R}}{(1+2b\widetilde{R})p^{0}}\right]\partial_{p^{a}}f=0.\label{eq: Vlasov 2}\end{equation}

Now, let us recall that two quantities are important for the Vlasov
equation in a curved spacetime \cite{key-18,key-22,key-26,key-28}.
The first is the current density

\begin{equation}
N^{\mu}=-\int\frac{dp}{p^{0}}\sqrt{g}p^{\mu}f\label{eq: corrente}\end{equation}

and the second is the stress-energy tensor

\begin{equation}
T^{\mu\nu}=-\int\frac{dp}{p^{0}}\sqrt{g}p^{\mu}p^{\nu}f.\label{eq: energia-impulso}\end{equation}

Here $g$ is the usual determinant of the metric tensor, which in
the case of the line-element (\ref{eq: metrica puramente scalare})
is given by

\begin{equation}
g=1+4b\widetilde{R}.\label{eq: determinante}\end{equation}

Keep in mind that both $N^{\mu}$ and $T^{\mu\nu}$ are divergence
free (conservation of energy): 

\begin{equation}
\begin{array}{c}
\nabla_{\mu}N^{\mu}=0,\\
\\\nabla_{\mu}T^{\mu\nu}=0.\end{array}\label{eq: cons. en.}\end{equation}

The mass shell conditions (\ref{eq: mass-shell}) can be rewritten
as 

\begin{equation}
p^{0}=\sqrt{(1+b\widetilde{R})^{-1}+\delta_{ab}p^{a}p^{b}.}\label{eq: mass-shell 2}\end{equation}

From the Christoffel connections (\ref{eq: connessioni}), computing
the Riemann tensor, Ricci tensor and Ricci scalar, the \emph{{}``effective}''
Einstein field equations 

\begin{equation}
G_{\mu\nu}=T_{\mu\nu},\label{eq: einstein 3}\end{equation}

can be obtained together with the \emph{{}``effective}'' Klein -
Gordon equation 

\begin{equation}
{}\square b\widetilde{R}=-T,\label{eq: KG 3}\end{equation}

where $T\equiv T_{\mu}^{\mu}$ is the trace of the stress-energy tensor.

To simplify the computations the analysis can be performed in a conformal
frame. Thus, rescaling the stress-energy tensor in the form 

\begin{equation}
T_{*}^{\mu\nu}=(1+3b\widetilde{R})T^{\mu\nu},\label{eq: riscalo 1}\end{equation}

one obtains

\begin{equation}
T_{*}=(1+3b\widetilde{R})T.\label{eq: riscalo 2}\end{equation}

Then, equation (\ref{eq: KG 3}) becomes

\begin{equation}
{}\square b\widetilde{R}=-T_{*}.\label{eq: KG 4}\end{equation}

It should be noticed that the particle density is still defined on
the mass shell of the starting line-element $g_{\alpha\beta}$. In
order to remove even this last connection with the starting frame
(Jordan Frame) we can rescale the momentum as 

\begin{equation}
p_{*}^{\mu}=(1+\frac{b\widetilde{R}}{2})p^{\mu}\label{eq: riscalo 3}\end{equation}

and we can define the particle density in the Einstein frame as

\begin{equation}
f_{*}(t,x,p_{*})=f\left(t,x,(1-\frac{b\widetilde{R}}{2})p_{*}\right).\label{eq: riscalo 4}\end{equation}

Hence, we can write our adaptation of the Vlasov system in the following
form 

\begin{equation}
{}\square b\widetilde{R}=(1+2b\widetilde{R})\int\frac{dp_{*}}{p_{*}^{0}}f_{*}(t,x,p_{*}),\label{eq: KG 5}\end{equation}

\begin{equation}
p_{*}^{0}=\sqrt{1+\delta_{ab}p_{*}^{a}p_{*}^{b}.}\label{eq: mass-shell 3}\end{equation}

\begin{equation}
\partial_{t}f_{*}+\frac{p_{*}^{a}}{p_{*}^{0}}\partial_{x^{a}}f_{*}-\frac{1}{p_{*}^{0}}[p_{*}^{\mu}\partial_{\mu}b\widetilde{R}p_{*}^{a}+\partial^{a}b\widetilde{R}]\partial_{p_{*}^{a}}f_{*}=0.\label{eq: Vlasov 3}\end{equation}

Because we want to restrict ourselves to spherical symmetry in the
present approximation, the line-element (\ref{eq: metrica puramente scalare})
can be rewritten as 

\begin{equation}
ds^{2}=[1+b\widetilde{R}(t,r)](dr^{2}-dt^{2}).\label{eq: metrica puramente scalare 2}\end{equation}

In this equation $r$ is the radial coordinate. Thus, in spherical
coordinates, equations (\ref{eq: KG 5}), (\ref{eq: mass-shell 3})
and (\ref{eq: Vlasov 3}) can be written as 

\begin{equation}
-\frac{d^{2}b\widetilde{R}}{dt^{2}}+\frac{1}{r^{2}}\frac{d}{dr}\left(\frac{d}{dr}b\widetilde{R}r^{2}\right)=(1+2b\widetilde{R})\mu(t,r),\label{eq: KG 6}\end{equation}

\begin{equation}
\mu(t,r)=\int\frac{dp}{\sqrt{1+p^{2}}}f(t,x,p),\label{eq: mass-shell 4}\end{equation}

\begin{equation}
\partial_{t}f+\frac{p}{\sqrt{1+p^{2}}}\cdot\partial_{x}f-\left[\left(\frac{d}{dt}b\widetilde{R}+\frac{x\cdot p}{\sqrt{1+p^{2}}}\frac{1}{r}\frac{d}{dr}b\widetilde{R}\right)p+\frac{x}{\sqrt{1+p^{2}}}\frac{1}{r}\frac{d}{dr}b\widetilde{R}\right]\cdot\partial_{p}f=0,\label{eq: Vlasov 4}\end{equation}

where the suffix $*$ has been removed for the sake of simplicity,
and we have denoted by $p$ the vector $p=(p_{1},p_{2},p_{3})$ with
$p^{2}=|p|^{2},$ and also defined $x$ for the vector $x_{i}=(x_{1},x_{2},x_{3})$.

Then, in searching for stationary states, and following \cite{key-16},
we can call $\lambda$ the wavelength of the \emph{{}``galactic''
gravitational wave} (\ref{eq: perturbazione totale}), i.e. the characteristic
length of our gravitational perturbation, and assume that $\lambda\gg d,$
where $d$ is the galactic scale, i.e. $d\thicksim10^{5}$ light-years
\cite{key-25}. In this way, the gravitational wave is {}``\emph{frozen-in}''
with respect to the galactic scale. 

Thus, the system of equations which defines the stationary solutions
of eqs. (\ref{eq: KG 6}), (\ref{eq: mass-shell 4}) and (\ref{eq: Vlasov 4}),
for our galaxy model is 

\begin{equation}
\frac{1}{r^{2}}\frac{d}{dr}\left(\frac{d}{dr}b\widetilde{R}r^{2}\right)=(1+2b\widetilde{R})\mu(r),\label{eq: KG 7}\end{equation}

\begin{equation}
\mu(r)=\int\frac{dp}{\sqrt{1+p^{2}}}f(x,p),\label{eq: mass-shell 5}\end{equation}

\begin{equation}
p\cdot\partial_{x}f-\frac{1}{r}\frac{d}{dr}b\widetilde{R}[(p\cdot x)p+x]\cdot\partial_{p}f=0,\label{eq: Vlasov 5}\end{equation}
In other words, the idea is that the spin-zero degree of freedom introduced
via the addition of an $R^{2}$ term in the gravitational Lagrangian
may be a candidate for the dark matter. This new degree of freedom
is termed the scalaron \cite{key-13}. 

Thus, we assume that within a galaxy the dominant contribution to
the curvature comes from the new degree of freedom (as it is also
the case to first approximation in the general relativity$+$cold
dark matter case). We deduce this contribution via the scalaron field
equation which itself has a baryonic source term. The baryons themselves
are then taken to evolve according to a collisionless Boltzmann equation,
propagating on a background perturbed by the scalaron. The collective
set of equations can in principle be solved for given initial data.

\section{The gravitational lensing }

The attentive reader notices that, in principle, as the metric is
conformally flat, this could mean that the scalaron Dark Matter in
galaxies will, in and of itself, result in no gravitational lensing
of light rays. This could be a serious problem for the model. This
is because there exists substantial evidence for gravitational lensing
by the large scale structure of a galaxy beyond that due to the constituent
baryons alone. For instance, the observed absence of lensing by the
effective \emph{dark matter} in a class of metric formulations of
\emph{Modified Newtonian Dynamics} formed the basis for a tentative
no-go theorem for such theories \cite{key-30}.

Actually, we show that in the proposed model the gravitational lensing
can be, in principle, obtained like an effect of spacetime curvature,
i.e. due by the Ricci curvature scalar.

In our linearized approach, gravitational lensing can be described
in a local Lorentz frame perturbed by the first order post-Newtonian
potential \cite{key-27}. Calling $V$ such a potential one can define
a refractive index \cite{key-27}

\begin{equation}
n\equiv1+2|V|.\label{eq: indice rifrazione}\end{equation}

Some clarifications are needed concerning this point. In the usual
\emph{Geometrical Optics}, the condition $n>1$ implies that the light
in a medium is slower than in vacuum \cite{key-36}. Then, the effective
speed of light in a gravitational field is expressed by \cite{key-36}

\begin{equation}
v=\frac{1}{n}\approx1-2|V|.\label{eq: vel mezzo}\end{equation}

Thus, one can obtain the Shapiro delay \cite{key-37} by integrating
over the optical path between the source and the observer:

\begin{equation}
\int_{source}^{observer}2|V|dl.\label{eq: sentiero}\end{equation}

The situation is analogous to the prism \cite{key-36}.

Now, we recall that, in the weak field approximation, the connection
between the post-Newtonian potential and the linearized theory is
given by the $g_{00}$ component of the line-element \cite{key-19}:

\begin{equation}
g_{00}\approx1+2V.\label{eq: connessione}\end{equation}

Thus, from eq. (\ref{eq: metrica puramente scalare}) one obtains

\begin{equation}
V\approx\frac{b\widetilde{R}(t,z)-1}{2},\label{eq: pot newt}\end{equation}

and 

\begin{equation}
n\approx1+2|b\widetilde{R}(t,z)-1|\label{eq: indice rifrazione 2}\end{equation}

is the equation which can, in principle, be used to discuss the gravitational
lensing in our model. Then, in this case, the gravitational lensing
is performed directly by spacetime curvature, i.e. by the Ricci scalar. 

For a sake of completeness, we discuss the gravitational lensing from
another point of view.

The condition $\lambda\gg d,$ on the characteristic length of the
gravitational perturbation, guarantees that the Ricci curvature remains
{}``\emph{frozen, i.e. constant},'' with respect to the galactic
characteristic distance scale. Thus, we put

\begin{equation}
b\widetilde{R}=K({\rm a\; constant}!).\label{eq: posto}\end{equation}
We can search deviations from standard General Relativity within the
galaxy by considering a spherically symmetric Schwarzschild-like metric
generated by the ordinary (barion) galaxy mass $M$ with the corrections
that are generated by curvature \cite{key-38}:

\begin{equation}
ds^{2}=-\exp(-\lambda r)dt^{2}+\exp(\lambda r)dr^{2}+r^{2}(d\theta^{2}+\sin^{2}\theta d\varphi^{2})\label{eq: Schwarzschild-like}\end{equation}

Then, following \cite{key-38}, the Ricci scalar is given by:

\begin{equation}
R=\exp(-\lambda r)(\lambda''-\lambda'^{2}+\frac{4\lambda'}{r}-\frac{2}{r^{2}})+\frac{2}{r^{2}},\label{eq: riccione}\end{equation}

where $'$ stands for the derivative with respect to $r.$

Putting the condition (\ref{eq: posto}) in Eq. (\ref{eq: riccione})
one gets \cite{key-38}:

\begin{equation}
\lambda(r)=-\ln\left(\frac{\alpha}{r}-\frac{\beta}{r^{2}}-\frac{K}{12b}r^{2}\right),\label{eq: landa}\end{equation}

and, by choosing $\alpha=-2M,$ $\beta=0$ in analogy with the standard
Schwarzschild metric, one gets \cite{key-19,key-38}

\begin{equation}
ds^{2}=-(1-\frac{2M}{r}-\frac{K}{12b}r^{2})dt^{2}+(1-\frac{2M}{r}+\frac{K}{12b}r^{2})^{-1}dr^{2}+r^{2}(d\theta^{2}+\sin^{2}\theta d\varphi^{2}).\label{eq: New Schwarzschild}\end{equation}

The physical interpretation implies that within some radius $r_{{\rm min}}$
deviation from General Relativity are highly suppressed and we say
that the object is \emph{screened} \cite{key-39,key-40,key-41}. This
is because we are assuming that, within the radius $r_{{\rm min}}$,
the spacetime curvature which is due by the \emph{ordinary}, i.e.
barion, mass of the galaxy dominates with respect to the \emph{intrinsic}
constant curvature $R=K$. On the other hand, for values $r>r_{{\rm min}}$,
the spacetime curvature which is due by the intrinsic curvature dominates
with respect to the spacetime curvature due by the barion mass of
the galaxy.

In order to clarify the issue of the bending of light in this $R^{2}$
scenario, in what follows, and as matter of illustration, we will
reproduce a piece of the discussion presented by Smith in Ref. \cite{key-41},
which, in comparison with our discussion at the beginning of Section
II, appears as a generalization of that digression (see the Appendix). 

Thus, for a general action $S=\int d^{4}x\sqrt{-g}(R+f(R)+\mathcal{L}_{m})$
screening occurs within a radius $r_{{\rm min}}$ implicitly given
by \cite{key-39,key-40,key-41}

\begin{equation}
|f'(R_{{\rm max}})|<\rho(r_{{\rm min}})r_{{\rm min}}^{2},\label{eq: screen 1}\end{equation}

where $\rho$ is the local value of the density.

In our case it is $f'(R)=2bR$ and $R_{{\rm max}}=\frac{K}{b}$. Then,
Eq. (\ref{eq: screen 1}) becomes

\begin{equation}
K<\frac{1}{2}\rho(r_{{\rm min}})r_{{\rm min}}^{2}.\label{eq: screen2}\end{equation}

Now, following \cite{key-41}, we can use Eq. (\ref{eq: screen2})
to further discuss the gravitational lensing.

Let us consider a galactic Navarro, Frank and White (NFW) halo of
the form \cite{key-40,key-41}

\begin{equation}
\rho(r)=\rho_{c}\delta_{c}\frac{r_{s}^{3}}{r(r+r_{s})^{2}},\label{eq: halo}\end{equation}

where $r_{s}$ is the scale radius and $\rho_{c}$ the critical density
of the universe \cite{key-40,key-41}.

By assuming the mass-concentration relation \cite{key-41,key-42,key-43}

\begin{equation}
c=\frac{9}{1+z}\left(\frac{M}{8.12\times10^{12}h^{-1}M_{\circledcirc}}\right),\label{eq: concentrazione}\end{equation}

where $z$ is the redshift, $M_{\circledcirc}$ the mass of the Sun
and $h$ is the Hubble parameter in units of $100km/(sMpc)$, the
amplitude $\delta_{c}$, that relates the concentration to the virial
radius with an overdensity $\triangle=119$ is given by \cite{key-41,key-42,key-43} 

\begin{equation}
\delta_{c}=\frac{\triangle\; c^{3}}{3[\ln(1+c)-(\frac{c}{1+c})]}.\label{eq: amplitude}\end{equation}

Eq. (\ref{eq: concentrazione}) reduces the NFW halo profile to a
one-parameter family which is taken to be dependent on the virial
mass, M \cite{key-41}. For $r<r_{s}$ $\rho_{NFW}$ scales like $r^{-1}$,
while for $r>r_{s}$ $\rho_{NFW}$ scales like $r^{-4}$ \cite{key-41}.
Thus, the innermost point at which deviations from general relativity
are suppressed in will occur at the scale radius $r_{s}$. Considering
Eq. (\ref{eq: screen2}) togheter with the NFW density profile on
sees that halos with masses which satisfy 

\begin{equation}
K>\rho_{c}\delta_{c}(M)\label{eq: screen 3}\end{equation}

As $\delta_{c}(M)$ decreases with decreasing $M$ \cite{key-41},
an upper limit is setted to the mass of halos which can be screened
given $K,$ i.e. given the value of the Ricci curvature, see Eq. (\ref{eq: posto}).
For masses below this threshold the theory strongly deviates from
general relativity and strong lensing around these halos is permitted.
We plan to further develop in this direction of research.

\section{Can $R^{2}$ gravity explain the Bullet Cluster dynamics?}

The scalaron has two aspects as a field (like classical electric or
magnetic field) and a particle (like quantized photon). We treated
the scalaron as a field in the Vlasov equation (see \cite{key-44}
for a recent cosmological application) whose scenario is in some sense
similar to that of the MOND \cite{key-45}. In such a scenario, it
seems to be difficult to explain the Bullet Cluster \cite{key-46}
although the Bullet Cluster could be naturally explained if the scalaron
could be a heavy particle . 

Aside from this, we emphasize that the main goal of this paper is
to give a potential solution to the Dark Matter Problem within galaxies.
Therefore, we do not ban the presence of a Dark Matter component in
our Universe which can explain properties of clusters of galaxies
which cannot be achieved by current alternative gravity theories \cite{key-46}.
In any case, in this Section we show that an important property of
the Bullet Cluster can in principle be explained in the scenario proposed
in this paper.

\subsection{A note on the Bullet Cluster 1E0657-56}

Since 2006, the argument ''But the Bullet Cluster ...'' has come into
scene in most discussions about the Dark Matter Problem in astrophysics.
It relates to the observations of the post-collision clusters of galaxies
in the source 1E0657-56 (known as the Bullet Cluster) \cite{key-46}.
It is seen that the X-ray emission is a mid the couple of colliding
clusters, and thus is shifted with respect to the bulk of matter making
up each group of galaxies, which resides in the outskirts. In most
studies of X-ray emission from galaxy clusters the gas usually sits
at its center. This offset can be understood by recalling that as
the gas pass each other it interacts electromagnetically, heats up,
ionize to emit X-rays, and hence is slowed-down during the collision.
This may explain why the gas appears a bit behind in the heading direction
of each of the group of galaxies involved. Meanwhile, non interacting
Dark Matter components do only interact through gravity and thus pass
each other unhindered without being slowed-down like the gas. In the
image, the field purported to represent the vast of mass in each cluster,
as compared to the X-ray emitting counterpart, is inferred from gravitational
lensing analysis of images of background galaxies, under the assumption
that general relativity is the theory of gravity (but be aware that
most modified gravity theories describe correctly the phenomenon of
gravitational lensing. See the Appendix). It follows the distribution
of galaxies, not the gas. As the most massive component of the galaxy
cluster is not centered at the gas, it is argued, upon that lens analysis,
that the large part of matter in the cluster is close to, and around
each galaxy group. And as the visible mass in the galaxies is not
enough to account for the velocity dispersion of a galaxy cluster,
it is suggested that there is Dark Matter, too. As noninteracting
component, the distribution of the Dark Matter can pass through each
other just like the galaxies, and that is why the bulk of mass should
be found close to the galaxies in a cluster collision like this. 

The analysis of the overimposed optical and X-ray images of the cluster
can be interpreted as an endorsement on the existence of Dark Matter.
This issue is based on a number of assumptions which are not totally
accepted by the astrophysics community because the large set of problems
for the Dark Matter hypothesis are, in any case, independent from
the observations of the Bullet Cluster. In a different context, the
collision velocity of the Bullet Cluster seems to be in disagreement
with the standard concordance model of cosmology \cite{key-48} (the
relative velocity of the clusters in the collision is an issue being
addressed next in the framework of this gravity theory). At the same
time, alternative theories of gravitation, while often said to fail
in explaining dynamics of galaxy clusters, can account for them rather
naturally. 

On the other hand, even if the Bullet Cluster can only be explained
by invoking Dark Matter, the problems on small scales persist. The
Bullet Cluster does not improve our understanding of the Local Group
of Galaxies, nor on the dynamics of individual galaxies, (as is on
focus in the present paper). The two arguments are absolutely independent.
A conclusion could be that the lensing mass and the hot gas have a
spatial shift, and hence the visible, hot gas cannot make up the bulk
of mass in the system. In any case, whether the cause of this phenomenon
is a nonstandard law of gravitation or missing mass is not that easy
to unveil.

\subsection{Scalaron and the Bullet Cluster dynamics}

As pointed out above, it is a common opinion that alternative gravity
theories cannot explain all the properties of the bullet cluster \cite{key-46},
even if scalar-vector-tensor theory could, in principle, take into
account the apparent discrepancy between the gravitational lensing
and X-ray maps of the colliding clusters, without requiring collisionless
Dark Matter \cite{key-47,key-50}. 

Meanwhile, hydrodynamic simulations of the colliding galaxy clusters
1E0657-06 show that at a distance of 4.6 Mpc between the two clusters,
an infall velocity of 3000 km/s is required in order to explain the
observed X-ray brightness and morphology of the cluster \cite{key-47,key-48}.
It has also been argued that such a high infall velocity is incompatible
with predictions of the standard cosmology (the concordance $\Lambda$CDM
model) \cite{key-47,key-49}.

Next, we extend our scalaron model in order to show that the infall
velocity of the two clusters can in principle be explained in the
framework introduced in the present paper. To rigorously show that
the scalaron gravity can cope with the Bullet Cluster dynamics we
need first to modify our assumptions. In particular, we will assume
that the characteristic length of the scalaron gravitational perturbation
is much longer not only than the galactic scale, but than the distance
between the colliding clusters too. In other words, the gravitational
wave is {}``\emph{frozen-in}'' with respect to the Bullet Cluster's
scale. In this way, we can consider the colliding clusters (the main
cluster and the {}``bullet'') like test particles in the gravitational
field of the scalaron. 

In the linearized approach of this paper, the coordinate system in
which the space-time is locally flat has to be used and the distance
between any two points (the colliding clusters) is given simply by
the difference in their coordinates in the sense of Newtonian physics
\foreignlanguage{italian}{\cite{key-19}}. This frame is the proper
reference frame of a local observer, which we assume to be located
within the main cluster. In this frame gravitational signals manifest
them-self by exerting tidal forces on the test masses. In other words,
we assume that the space-time within the main cluster is locally flat
with respect to the global curvature generated by the scalaron which
is described by the line element (\ref{eq: metrica puramente scalare}).
By using the proper reference frame of a local observer the time coordinate
$x_{0}$ is the proper time of the observer O and the spatial axes
are centred in O. In the special case of zero acceleration and zero
rotation the spatial coordinates $x_{j}$ are the proper distances
along the axes and the frame of the local observer reduces to a local
Lorentz frame \foreignlanguage{italian}{\cite{key-19}}. As the origin
of the coordinates is located within the main cluster, the proper
distances between the two clusters are the coordinates of the {}``bullet''.
The line element is \foreignlanguage{italian}{\cite{key-19}}

\begin{equation}
ds^{2}=+(dx^{0})^{2}-\delta dx^{i}dx^{j}-O(|x^{j}|^{2})dx^{\alpha}dx^{\beta}.\label{eq: metrica local lorentz}\end{equation}

The effect of the gravitational force on test masses is described
by the equation

\begin{equation}
\ddot{x^{i}}=-\widetilde{R}_{0k0}^{i}x^{k},\label{eq: deviazione geodetiche}\end{equation}
which is the equation for geodesic deviation in this frame \foreignlanguage{italian}{\cite{key-19}}.
$\widetilde{R}_{0k0}^{i}$ is the linearized Riemann tensor\foreignlanguage{italian}{
and $x^{i}$ }are the coordinates of the {}``bullet'' which represent\foreignlanguage{italian}{
the separation vector between the two test masses \cite{key-19}.}

To study the effect of the scalaron on the two clusters, $\widetilde{R}_{0k0}^{i}$
has to be computed in the proper reference frame of the main cluster.
But, as the linearized Riemann tensor $\widetilde{R}_{\mu\nu\rho\sigma}$
is invariant under gauge transformations \foreignlanguage{italian}{\cite{key-19}},
it can be directly computed from Eq. (\ref{eq: metrica puramente scalare}). 

From \foreignlanguage{italian}{\cite{key-19}} we get

\begin{equation}
\widetilde{R}_{\mu\nu\alpha\beta}=\frac{1}{2}\{\partial_{\mu}\partial_{\beta}h_{\alpha\nu}+\partial_{\nu}\partial_{\alpha}h_{\mu\beta}-\partial_{\alpha}\partial_{\beta}h_{\mu\nu}-\partial_{\mu}\partial_{\nu}h_{\alpha\beta}\},\label{eq: riemann lineare}\end{equation}

that, in the case Eq. (\ref{eq: metrica puramente scalare}), gives

\begin{equation}
\widetilde{R}_{0\gamma0}^{\alpha}=\frac{b}{2}\{\partial^{\alpha}\partial_{0}\widetilde{R}\eta_{0\gamma}+\partial_{0}\partial_{\gamma}\widetilde{R}\delta_{0}^{\alpha}-\partial^{\alpha}\partial_{\gamma}\widetilde{R}\eta_{00}-\partial_{0}\partial_{0}\widetilde{R}\delta_{\gamma}^{\alpha}\}.\label{eq: riemann lin scalare}\end{equation}

The different elements are (only the non zero ones will be written
down explicitly) 

\begin{equation}
\partial^{\alpha}\partial_{0}\widetilde{R}\eta_{0\gamma}=\left\{ \begin{array}{ccc}
\partial_{t}^{2}\widetilde{R} & for & \alpha=\gamma=0\\
\\-\partial_{z}\partial_{t}\widetilde{R} & for & \alpha=3;\gamma=0\end{array}\right\} \label{eq: calcoli}\end{equation}

\begin{equation}
\partial_{0}\partial_{\gamma}\widetilde{R}\delta_{0}^{\alpha}=\left\{ \begin{array}{ccc}
\partial_{t}^{2}\widetilde{R} & for & \alpha=\gamma=0\\
\\\partial_{t}\partial_{z}\widetilde{R} & for & \alpha=0;\gamma=3\end{array}\right\} \label{eq: calcoli2}\end{equation}

\begin{equation}
-\partial^{\alpha}\partial_{\gamma}\widetilde{R}\eta_{00}=\partial^{\alpha}\partial_{\gamma}\widetilde{R}=\left\{ \begin{array}{ccc}
-\partial_{t}^{2}\widetilde{R} & for & \alpha=\gamma=0\\
\\\partial_{z}^{2}\widetilde{R} & for & \alpha=\gamma=3\\
\\-\partial_{t}\partial_{z}\widetilde{R} & for & \alpha=0;\gamma=3\\
\\\partial_{z}\partial_{t}\widetilde{R} & for & \alpha=3;\gamma=0\end{array}\right\} \label{eq: calcoli3}\end{equation}

\begin{equation}
-\partial_{0}\partial_{0}\widetilde{R}\delta_{\gamma}^{\alpha}=\begin{array}{ccc}
-\partial_{t}^{2}\widetilde{R} & for & \alpha=\gamma\end{array}.\label{eq: calcoli4}\end{equation}

By inserting these results in Eq. (\ref{eq: riemann lin scalare})
we obtain 

\begin{equation}
\begin{array}{c}
\widetilde{R}_{010}^{1}=-\frac{b}{2}\ddot{\widetilde{R}}\\
\\\widetilde{R}_{010}^{2}=-\frac{b}{2}\ddot{\widetilde{R}}\\
\\\widetilde{R}_{030}^{3}=\frac{b}{2}(\partial_{z}^{2}\widetilde{R}-\partial_{t}^{2}\widetilde{R}).\end{array}\label{eq: componenti riemann}\end{equation}

At the Bullet Cluster's scale we have homogeneity and isotropy, which
imply $\partial_{z}^{2}\widetilde{R}=0$. Therefore, Eqs. (\ref{eq: componenti riemann})
become

\begin{equation}
\begin{array}{c}
\widetilde{R}_{010}^{1}=-\frac{b}{2}\ddot{\widetilde{R}}\\
\\\widetilde{R}_{010}^{2}=-\frac{b}{2}\ddot{\widetilde{R}}\\
\\\widetilde{R}_{030}^{3}=-\frac{b}{2}\ddot{\widetilde{R}}.\end{array}\label{eq: componenti riemann 2}\end{equation}

By using Eq. (\ref{eq: deviazione geodetiche}), we obtain

\begin{equation}
\ddot{x}=\frac{b}{2}\ddot{\widetilde{R}}x,\label{eq: accelerazione mareale lungo x}\end{equation}

\begin{equation}
\ddot{y}=\frac{b}{2}\ddot{\widetilde{R}}y\label{eq: accelerazione mareale lungo y}\end{equation}

and 

\begin{equation}
\ddot{z}=\frac{b}{2}\ddot{\widetilde{R}}z.\label{eq: accelerazione mareale lungo z}\end{equation}

These are three perfectly symmetric oscillations of the scalaron (wave-packet).
If one re-introduces the radial distance $r$, Eqs. (\ref{eq: accelerazione mareale lungo x}),
(\ref{eq: accelerazione mareale lungo y}) and (\ref{eq: accelerazione mareale lungo z})
are summarized by

\begin{equation}
\ddot{r}=\frac{b}{2}\ddot{\widetilde{R}}r.\label{eq: accelerazione mareale radiale}\end{equation}

Eq. (\ref{eq: accelerazione mareale radiale}) can be solved with
the perturbation method \cite{key-19}

\begin{equation}
r(t)\backsimeq r(0)[1+\frac{b}{2}\widetilde{R}(t)].\label{eq: spostamento lungo r}\end{equation}

Assuming that the wave-packet is in the contraction phase we get the
relative infall velocity as

\begin{equation}
v_{infall}=\frac{d}{dt}r(t)\backsimeq r(0)\frac{b}{2}\dot{\widetilde{R}}(t).\label{eq: v infall}\end{equation}

Therefore, in order to have consistence with the observations we need

\begin{equation}
r(0)\frac{b}{2}\dot{\widetilde{R}}(t)\backsimeq3000\mbox{ }km/s.\label{eq: pre-consistenza}\end{equation}
Being $r(0)\backsimeq4.6\mbox{ }Mpc$, the model will be consistent
with observations if

\begin{equation}
\frac{b}{2}\dot{\widetilde{R}}(t)\backsimeq\frac{3000\mbox{ }km}{4.6\mbox{ }Mpc}s^{-1}\backsimeq2*10^{-17}s^{-1}.\label{eq: consistenza}\end{equation}

The assumption that the characteristic length of the scalaron gravitational
perturbation is much longer than the distance between the colliding
clusters implies that the frequency of the wave-packet has to be $\omega_{E}\ll10^{-14}s^{-1}$.
For example, for a value $\omega_{E}\sim10^{-16}s^{-1}$ Eq. (\ref{eq: consistenza})
guarantees theoretical consistence if $\frac{b}{2}\widetilde{R}\sim10^{-1}.$

\section{Final discussion and closing remarks}

It has been shown that in the framework of $\textbf{\ensuremath{R^{2}}}$gravity
and in the linearized approach, it is possible to obtain spherically
symmetric and stationary galaxy states which can be interpreted like
an approximated solution of the Dark Matter problem. In fact, in the
model the Ricci curvature generates a high energy term that can in
principle be identified as the dark matter content of most the galaxies.
The model can also help to gain a better understanding of the basics
of the Einstein-Vlasov system. 

In other words, we have discussed, in the linearized $\textbf{\ensuremath{R^{2}}}$gravity,
the solutions of a Klein-Gordon equation for the spacetime curvature
like galactic high energy \emph{scalarons}, i.e. particles that has
been analysed like wave-packets, and we have shown stationary solutions
in terms of the Einstein-Vlasov system. 

In this approximation, the energy of the galaxy-particle has been
seen like the dark matter component which guarantees the equilibrium
of galaxies.

An important point is that, because of the high energy of such particle-galaxies,
the constant coupling of the $\textbf{\ensuremath{R^{2}}}$ in the
gravitational action comes out to be infinitesimal with respect to
the linear term $R\mbox{\mbox{. }}$ In this way, the deviation from
standard general relativity is very weak, and, in principle, the theory
could pass the Solar System tests. An analysis on the gravitational
lensing has been also performed in Section 4. 

We recall that the main goal of this paper is to give a potential
solution to the Dark Matter Problem within galaxies. Thus, we do not
ban the presence of a Dark Matter component in our Universe which
can explain properties of clusters of galaxies which cannot be achieved
by alternative gravity theories. In any case, in Section 5 we showed
that an important property of the bullet cluster, i.e. an infall velocity
of 3000 km/s between the two clusters, can be, in principle, explained
in the scenario given in this paper.

\section*{Acknowledgements }

The authors thank an unknown referee for precious advices and comments
which permitted to improve this paper. Christian Corda thanks the
R. M. Santilli Foundation for partially supporting this research (Research
Grant Number RMS-TH-5735A2310). Herman J. Mosquera Cuesta thanks Fundação
Cearense para o Desenvolvimento Cientìfico e Tecnológico (FUNCAP),
Ceará, Brazil for financial support.

\section*{Appendix: dynamics in $f(R)$ theories of gravity and gravitational
lensing }

We emphasize that hereafter we closely follow the paper by Smith \cite{key-41},
and the set of references quoted there. A generic branch of scalar-tensor
theories in which the Brans-Dicke parameter $\omega\equiv0$! bring
in unique predictions: they feature a peculiar effect known as the
chameleon mechanism, by which the modifications to general relativity
get quickly suppressed nearby sufficiently massive objects, such as
galaxies. This rapid decay in the theory's gravitational effects present
a unique size-dependent lensing footprint which is discussed next. 

The action for $f(R)$-theories (functional of the Ricci scalar $R$) 

\begin{equation}
S=\frac{1}{2\kappa}\int d^{4}x\sqrt{-g}\left[R+f(R)\right]+S_{m},\label{action}\end{equation}

where $S_{m}$ action for the matter fields. The field equations turn
out to be

\begin{eqnarray}
\left[1+f'(R)\right]G_{\mu\nu} & + & \frac{1}{2}g_{\mu\nu}\left[Rf'(R)-f+2\Box f'(R)\right]\nonumber \\
 &  & -\nabla_{\mu}\nabla_{\mu}f'(R)=\kappa T_{\mu\nu}.\label{eq:fR_fieldequation}\end{eqnarray}

In this class of theories the Ricci scalar becomes a dynamical quantity
whose equation of motion is determined by the trace of the field equation, 

\begin{equation}
\Box f'(R)=\frac{1}{3}\bigg(\kappa T+R\left[1-f'(R)\right]+2f\bigg).\label{eq:trace}\end{equation}

In the limit where $f\rightarrow0$, Eq. (\ref{eq:trace}) reduces
to the standard relation $R=-\kappa T$ of general relativity. Here
$T$ is the trace of the stress-energy tensor. 

Using the trace equation to rewrite the gravitational equation of
motion one obtains

\begin{equation}
T_{{\rm eff}}=\frac{1}{3\kappa}\left[\kappa T+R\right]+\frac{1}{3\kappa}\left[2Rf'(R)-f\right].\end{equation}

Solutions to the Eq. (\ref{eq:trace}) determine the lensing predictions
for the theory, and can be understood by rewriting the trace of the
field equation in the form

\begin{equation}
\Box f'(R)=-\frac{\mathrm{d}V}{\mathrm{d}f'(R)},\end{equation}

where

\begin{equation}
\frac{\mathrm{d}V}{\mathrm{d}f'(R)}\equiv\frac{1}{3}\bigg(\kappa T+R\left[1-f'(R)\right]+2f\bigg).\end{equation}

Notice that for functions $f(R)$ which reproduce the observed expansion
history of the universe, the minimum of this potential yields the
general relativistic relationship between $R$ and $T$, i.e, $R=-\kappa T$. 

Hence, within a given object, a typical galaxy; for instance, far
away from the center the scalar curvature starts off nearly at its
asymptotic value, $R_{{\rm max}}$, and evolves with radial distance
from the center. If the object is too small compared to the wavelength
of the gravitational perturbation, then $R\sim R_{{\rm max}}$ throughout
the object, and deviations from general relativity will be of order
unity. Meanwhile, if the object is large enough then the scalar curvature
is forced to the minimum of its potential and $R=-\kappa T$ within
some radius $r_{{\rm min}}$. Within that radius deviations from general
relativity are highly suppressed, and thus one says that the object
is \emph{{}``screened}''. Finally, on the generic prospective for
it to give rise to dark matter within extended gravity, we recall
that a further argument in favor of the potential of modifying the
gravity theory at the galactic scale so as to explain dark matter
has been provided by Ellis \cite{key-24}.

\end{document}